\documentstyle[11pt,newpasp,twoside,epsf]{article}
\begin{document}
\title{Criteria for spectral classification of cool stars in the near-IR GAIA
        wavelength region}
 \author{F. Boschi and U. Munari}
\affil{Padova Astronomical Observatory -- INAF, Asiago Station, I-36012 Asiago (VI), Italy}
 \author{R. Sordo and P.M. Marrese}
\affil{Asiago Astrophysical Observatory of the Department of Astronomy, Univ. of Padova, I-36012 Asiago (VI), Italy}

\begin{abstract}
The far-red portion of the spectrum offers bright prospects for an accurate
classification of cool stars, like the giant components of symbiotic stars.
The 8480--8740 \AA\ region, free from telluric absorptions and where the
GAIA Cornerstone mission by ESA will record spectra for $3\times 10^8$
stars, is investigated on the base of available observed and synthetic
spectral atlases. We have identified and calibrated diagnostic line ratios
useful to derive the effective temperature (spectral type) and gravity
(luminosity class) for cool stars observed at spectral resolutions 10,000
$\leq \lambda/\Delta\lambda \leq$ 20,000, bracketing that eventually 
chosen for GAIA.
\end{abstract}

\section{Introduction}

The GAIA Cornerstone mission by ESA (Perryman et al. 2001), scheduled for
launch around 2010, will record during its 5-year lifetime an average of
$\sim$100 spectra for each of all the stars brighter than $V\sim$17.5 mag
($\sim 3 \times 10^8$ objects), over the wavelength range 8480--8740 \AA\
and at a resolution 10\,000$ \la \lambda/\Delta\lambda \la$ 20\,000,
corresponding to dispersions 0.22 $\la$ \AA/pix $\la$ 0.44 in the Nyquist
FWHM=2~pix sense. The main aim of such spectra is to support the mission
micro-arcsec accurate astrometry with the 6$^{th}$ component of the
phase-space via determination of the radial velocities. However, this huge
number of spectra will not only provide radial velocities but will also
carry the whole usual astrophysical content (Munari 2002), which will be
pretty large given the adopted high spectral resolution and the diagnostic
potential of this wavelength region dominated by the CaII triplet, the head
of the Paschen series, multiplet \#1 and \#8 of NI, and a forest of FeI, TiI
lines and of many other metals.

The study of the cool component of symbiotic stars will particularly
benefit from the choice of this wavelength region for the GAIA mission, as
much as the spectral observations performed from the ground (cf. Marrese et
al., this volume). In fact, the 8480--8740 \AA\ region is the only
astrophysically relevant wavelength interval longward of H$\alpha$ free from
telluric absorption interference (cf. Munari 1999), and is within the
reach of many of the currently available high-resolution spectrographs.

\begin{figure}
\plotone{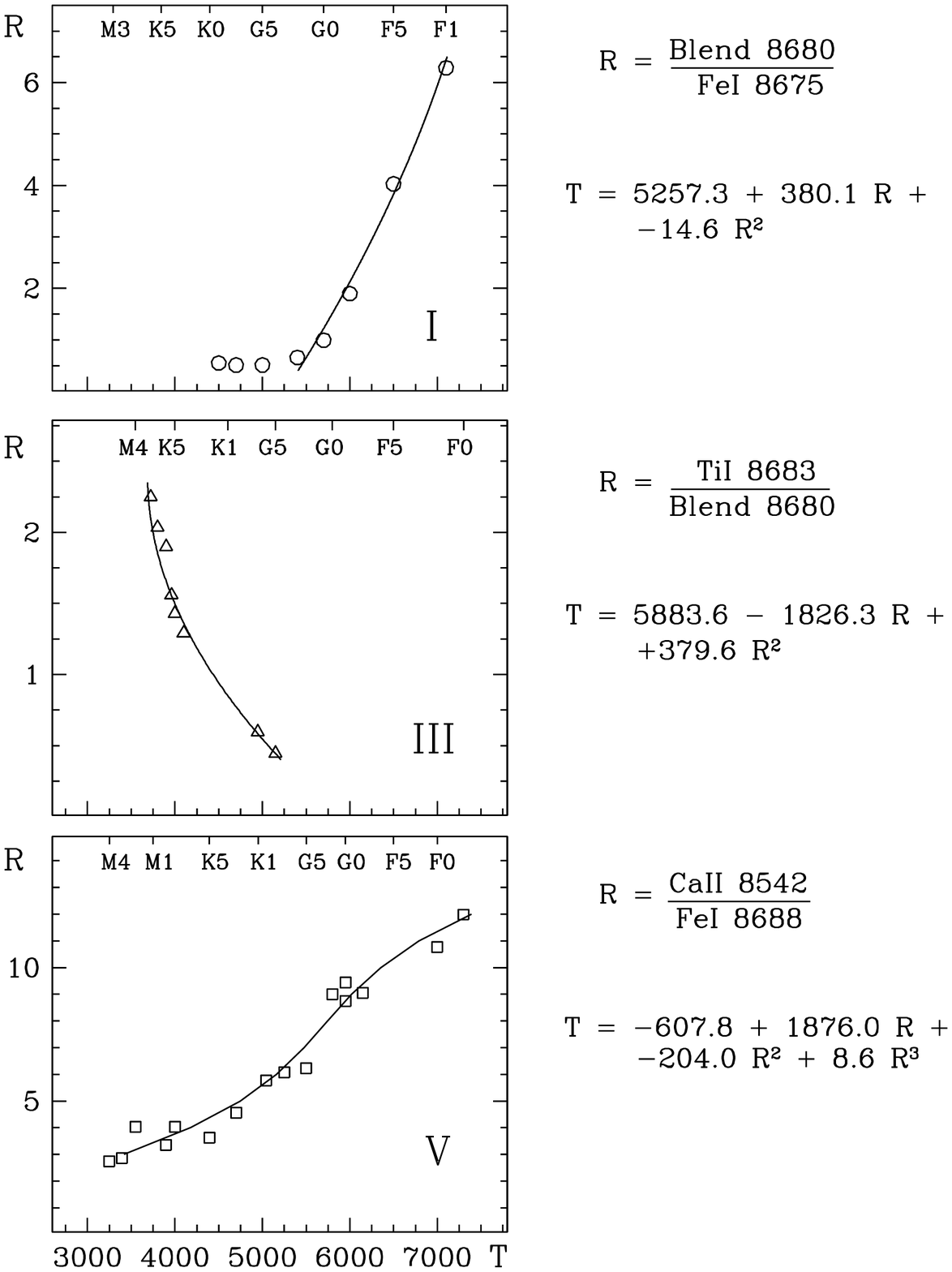}
\caption{Examples of temperature (spectral type) diagnostic line 
ratios over the GAIA wavelength range.}
\end{figure}

\begin{figure}
\plotone{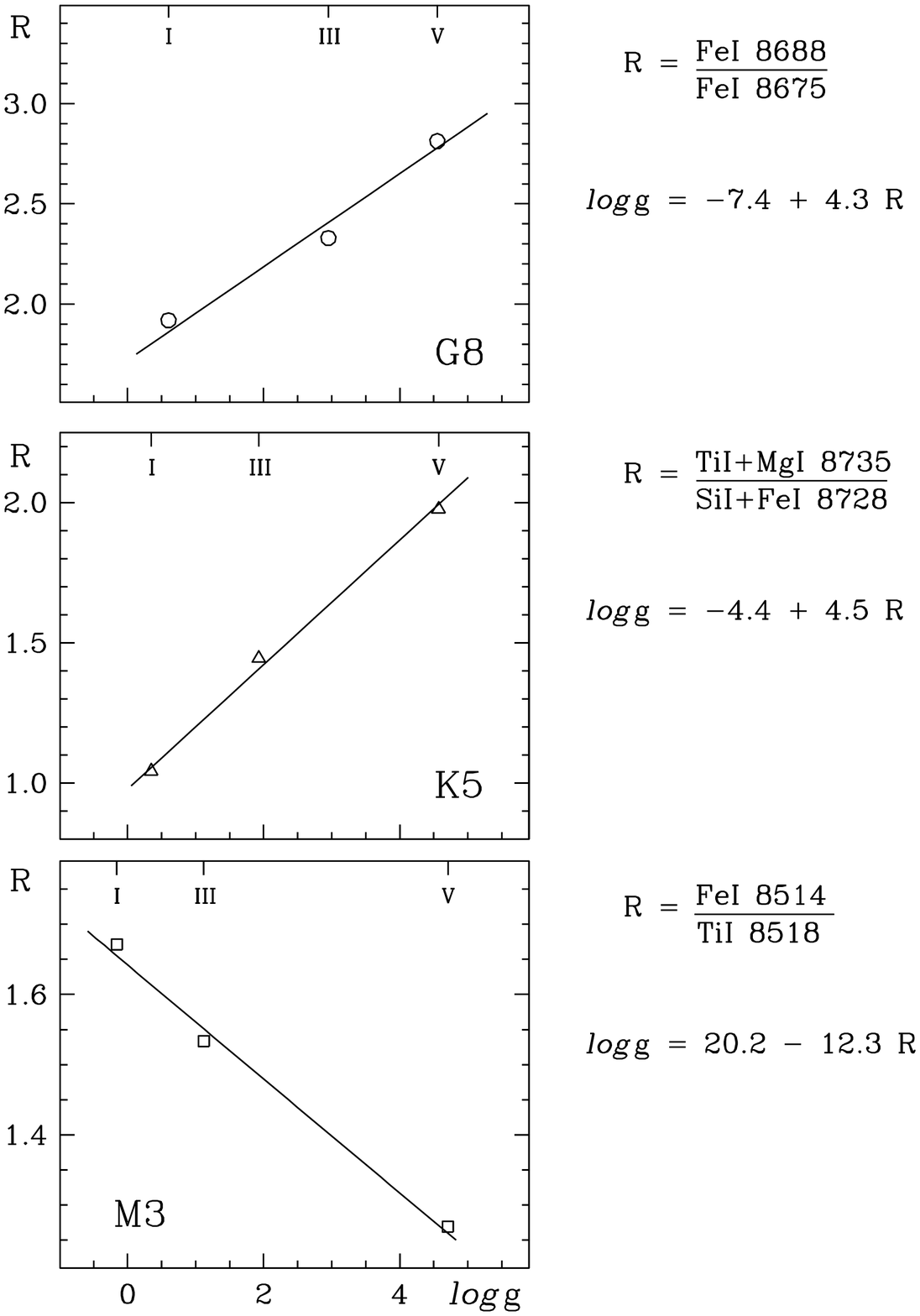}
\caption{Examples of gravity (luminosity class) diagnostic line 
ratios over the GAIA wavelength range.}
\end{figure}

We have identified and calibrated, over this wavelength interval, a family
of diagnostic line ratios that provide good estimate of temperature (spectral
type) and surface gravity (luminosity). The whole set of ratios will be
described elsewhere, with only a sample of them presented here. Other
diagnostic ratios have been presented by Munari (2002).

The line ratios have been investigated on the available high-resolution
spectral atlases in this region, both observational (Munari and Tomasella
1999) and synthetic (Munari and Castelli 2000, Castelli and Munari 2001)
ones. The relations here presented come entirely from observational
material, mainly referring to MKK standards (thus bright, solar neighborhood
stars).

Figure~1 presents temperature (spectral type) sensitive indicators for
supergiant, giant and dwarf cool stars, built on the ratio of equivalent
widths. To {\em Blend 8680} contribute mainly lines of FeI (8679.638), SI
(8678.927, 8679.620, 8680.411) and NI (8680.282), and the extremes for
equivalent width integration extends from 8678.0 to 8681.6 \AA. The extremes
for CaII~8542 extend from 8522.0 to 8565.0 \AA, to fully cover the line
wings (even if including several weak metallic lines). Those for FeI~8675 go
from 8673.0 to 8676.5, for TiI~8683 from 8681.6 to 8684.0, for FeI~8688 from
8688.5 to 8690.5 \AA.

Figure~2 offers surface gravity (luminosity class) sensitive indicators for
supergiant, giant and dwarf cool stars, built on the ratio of equivalent
widths. The extremes for equivalent width integration of FeI~8688 and
FeI~8675 are the same for in Figure~1, while for TiI+MgI~8735 it is
8734.3--8737.7, for SiI+FeI~8728 is 8726.7--8729.5, for FeI~8514 is
8512.5--8515.8 and for TiI~8518 it is 8516.6--8519.6 \AA.


\begin{references}
\reference Castelli F. \& Munari U. 2001, \aap, 366, 1003
\reference Munari U. 1999, Baltic Astron. 8, 73
\reference Munari U. 2002, in {\em GAIA: An European Space Project}, Les Houches, O.Bienaym\'e and C.Turon ed.s,
           EAS Pub. Series, EDP Sciences, pag. 39
\reference Munari U. \& Castelli F. 2000, \aaps, 141, 141
\reference Munari U. \& Tomasella L. 1999, \aaps, 137, 521
\reference Perryman M.A.C. et al. 2001, \aap, 369, 339
\end{references}
\end{document}